\documentclass[aoas]{imsart}

\RequirePackage{amsthm,amsmath,amsfonts,amssymb}
\RequirePackage[authoryear]{natbib}
\RequirePackage[colorlinks,citecolor=blue,urlcolor=blue]{hyperref}
\RequirePackage{graphicx}
\RequirePackage{booktabs,multirow}
\RequirePackage{float}

\startlocaldefs

\endlocaldefs

\begin{document}

\begin{frontmatter}
\title{Walking fingerprinting using wrist accelerometry during activities of daily living in NHANES}
\runtitle{NHANES fingerprinting}

\begin{aug}
\author[A]{\fnms{Lily}~\snm{Koffman}\ead[label=e1]{lkoffma2@jh.edu}\orcid{0000-0003-1543-2896}},
\author[A]{\fnms{John}~\snm{Muschelli}~III\orcid{0000-0001-6469-1750}},
\and
\author[A]{\fnms{Ciprian}~\snm{Crainiceanu}\orcid{0000-0001-6601-3881}}

\address[A]{Department of Biostatistics, Johns Hopkins Bloomberg School of Public Health\printead[presep={,\ }]{e1}}
\end{aug}

\begin{abstract}
We propose a method for identifying individuals based on their continuously monitored wrist-worn accelerometry during activities of daily living.
The method consists of three steps: (1) using Adaptive Empirical Pattern Transformation (ADEPT), a highly specific method to identify walking; (2) transforming the accelerometry time series into an image that corresponds to the joint distribution of the time series and its lags; and (3) using the resulting images to construct a person-specific walking fingerprint. The method is applied to $15{,}000$ individuals from the National Health and Nutrition Examination Survey (NHANES) with up to $7$ days of wrist accelerometry data collected at $80$ Hertz. The resulting dataset contains more than $10$ terabytes, is roughly $2$ to $3$ orders of magnitude larger than previous datasets used for activity recognition, is collected in the free living environment, and does not contain labels for walking periods. Using extensive cross-validation studies, we show that our method is highly predictive and can be successfully extended to a large, heterogeneous sample representative of the U.S. population: in the highest-performing model, the correct participant is in the top 1\% of predictions 96\% of the time. 
 
\end{abstract}


\end{frontmatter}


\section{Introduction}
\label{sec:intro}
\subsection{Background}\label{subsec:background}
People can be identified by their walking pattern/gait. This has been shown in small sample sizes with data collected in lab or semi-controlled environments. Published studies have focused on person identification using data from video, underfoot force sensors, or inertial devices placed on the legs or torso \citep{connor_biometric_2018}. These studies provide crucial information about what is possible in ideal in-lab conditions but have not been deployed in real world scenarios. These approaches place a high burden on study participants and are often computationally expensive, which make them impractical at scale \citep{gait_survey, gait_survey2}. Therefore, recent developments have focused on gait-based identification using wrist-worn accelerometers \citep{connor_biometric_2018}, which provide convenient, unobtrusive, long term-data collection in large populations for extended periods of time \citep{karas_accelerometry_2019}. Participant identification from accelerometry data is useful because it forms the foundation of walking as a biomarker. If individuals have a walking-specific ``fingerprint,'' i.e. a unique walking pattern, then changes in this pattern could be indicative of changes in health status, such as increased fall risk or early onset of disease. The walking fingerprint could also be used as a metric for effectiveness of an intervention or recovery from a surgery or injury.

Several methods have been developed for gait-based identification from wrist-worn accelerometry data including: (1) matching based on features of the step cycle (the period of a step from heel strike through to toe-off) \citep{mantyjarvi_identifying_2005, derawi_unobtrusive_2010, gafurov_biometric_2006, gafurov_improved_2010, derawi_unobtrusive_2010, rong_wearable_2007}; (2) hidden Markov models trained on the raw triaxial acceleration time series \citep{hmm}; (3) clustering or voting based on signature points \citep{signature_points, zju_data}; and (4) our own walking fingerprinting based on the empirical joint distribution of the acceleration and lag acceleration \citep{Koffman2023, Koffman2024}. These methods have been applied in smaller datasets ($\leq 50$ study participants, except \cite{zju_data}, which included $175$ study participants), where walking data was collected in controlled or semi-controlled settings and walking labels were available.  

In this paper we address the question of whether individuals can be identified from their high-resolution wrist accelerometry data during walking, without the availability of walking labels, in large heterogeneous studies of daily living.  This problem is inspired by the increased availability of high resolution wrist activity data collected on tens of thousands of study participants in the free living environment, including the Nutrition Examination Survey (NHANES) \citep{nhanes} and the UK Biobank \citep{UKB}. These massive datasets present new opportunities and challenges for gait-based identification from accelerometry. Opportunities include longer observation times of walking, realistic scenarios and activity contexts, and a large population of individuals. Challenges include the lack of walking labels, the computational difficulty of fitting models on thousands of individuals, and increased likelihood of having individuals with similar walking patterns. To our knowledge, no gait-based identification method has been applied in a free living accelerometry dataset. Furthermore, we are not aware of identification methods for any modality of gait data that have been deployed at scale in the free-living environment. This is a difficult problem: even facial recognition, a well-established biometric \citep{facial_recog}, is not perfect in large, free-living samples \citep{megaface}.

This paper deploys walking fingerprinting in free living accelerometry data collected from the NHANES dataset. It is different than previous work in two important ways: first, the accelerometry data are free-living and unlabeled, so we do not know when individuals are walking, and second, the sample size ($n > 15{,}000$) is two orders of magnitude larger than any previously used datasets. The free-living nature of the data require additional steps to identify walking, and the large sample size requires computational and modeling changes to the existing method.

\subsection{Data description}\label{subsec:datadescription}
NHANES is a large, nationally representative study of over $5{,}000$ individuals every two years. Data are collected on demographic, socioeconomic, and health-related information and are publicly available. In the NHANES 2011-2012 and 2013-2014 waves, participants were provided with a wrist-worn accelerometer on the day of their Mobile Examination Center visit. They were instructed to wear the ActiGraph GT3X+ (Pensacola, FL) continuously on the non-dominant wrist for seven consecutive days and return the accelerometer by mail on the morning of the ninth day \citep{paxming, paxminh}. The same protocol was used in the NHANES National Youth Fitness Survey (NNYFS), which was conducted in 2012 to collect data on physical activity of children in the U.S. ages $3$ to $15$ \citep{paxminy}. 
The raw, triaxial $80$ Hertz data were made available by NHANES in 2022. Accelerometer data was captured for $6{,}917$ individuals in NHANES 2011-2012, $7{,}776$ individuals in NHANES 2013-2014, and $1{,}477$ children in NNYFS. Compliance was high; $96$\% of participants with data wore the device until the ninth day and only $2$\% of participants wore the device for fewer than seven days \citep{paxming, paxminh, paxminy}. More details on the procedure for the accelerometers are available at  \url{https://wwwn.cdc.gov/Nchs/Data/Nhanes/Public/2011/DataFiles/PAXDAY_G.htm}.

\begin{figure}[H]
    \centering
    \includegraphics[width=\linewidth]{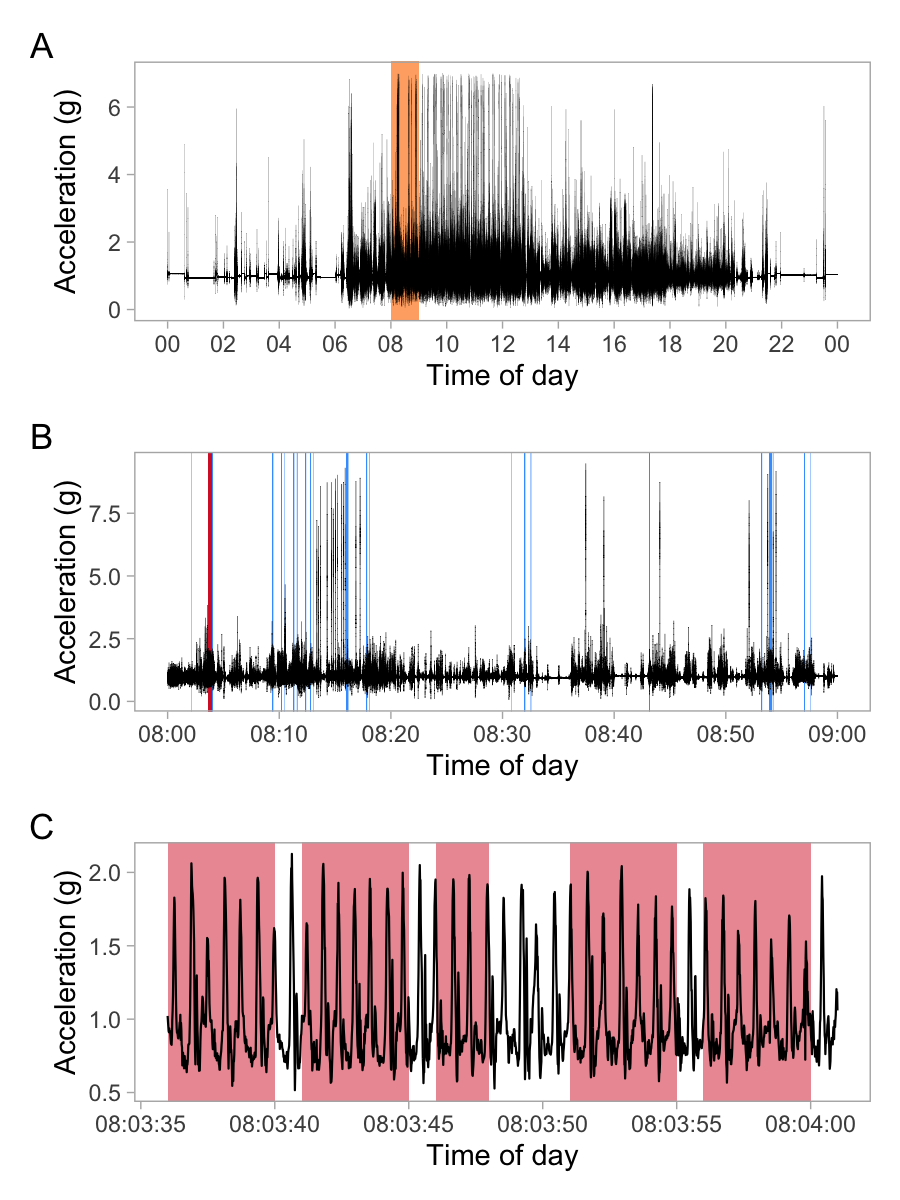}
    \caption{A sample of raw accelerometry data from one participant in NHANES. Panel A displays one full day of data, where observations are captured every 1/80th of a second. Panel B zooms in on one hour of the data (the area highlighted in orange in panel A); colored areas are areas ADEPT-identified walking. Panel C further zooms in on 30 seconds of data (the area highlighted as red in panel B). In Panel C, red highlighted areas denote ADEPT-identified walking.}
    \label{fig:rawacc}
\end{figure}
The raw data were downloaded from the NHANES website. All data were considered for this analysis (no wear time criteria were applied). Data were collected along three orthogonal axes, but for the purpose of this paper we work with the vector magnitude (square root of the sum of squared observations along the three axes), because vector magnitude is rotationally invariant and therefore not sensitive to movements of the accelerometer on the wrist. To demonstrate the data structure we plot one day of accelerometry from a single participant in Figure~\ref{fig:rawacc}. In panel A, data for the full day are shown at the sample rate of 80 observations per second. The hour from 8 to 9AM, which is highlighted in orange, is zoomed in on in panel B. In panel B, the colored bars indicate areas where walking is identified by Adaptive Empirical Pattern Transformation (ADEPT) \citep{adept}, a highly specific walking identification algorithm. In panel C, the red highlighted area from panel B is zoomed in on, and areas identified as walking by ADEPT are highlighted in red. While all of the seconds shown in panel C may be walking, ADEPT does not identify all of these seconds as walking. Our method does not require correct identification of all walking periods: it just requires that some walking periods are identified. These periods are used to build the walking fingerprint. 

\subsection{Statistical Challenges}\label{subsec:challenges}

\begin{figure}[ht]
    \centering
    \includegraphics[width=\linewidth]{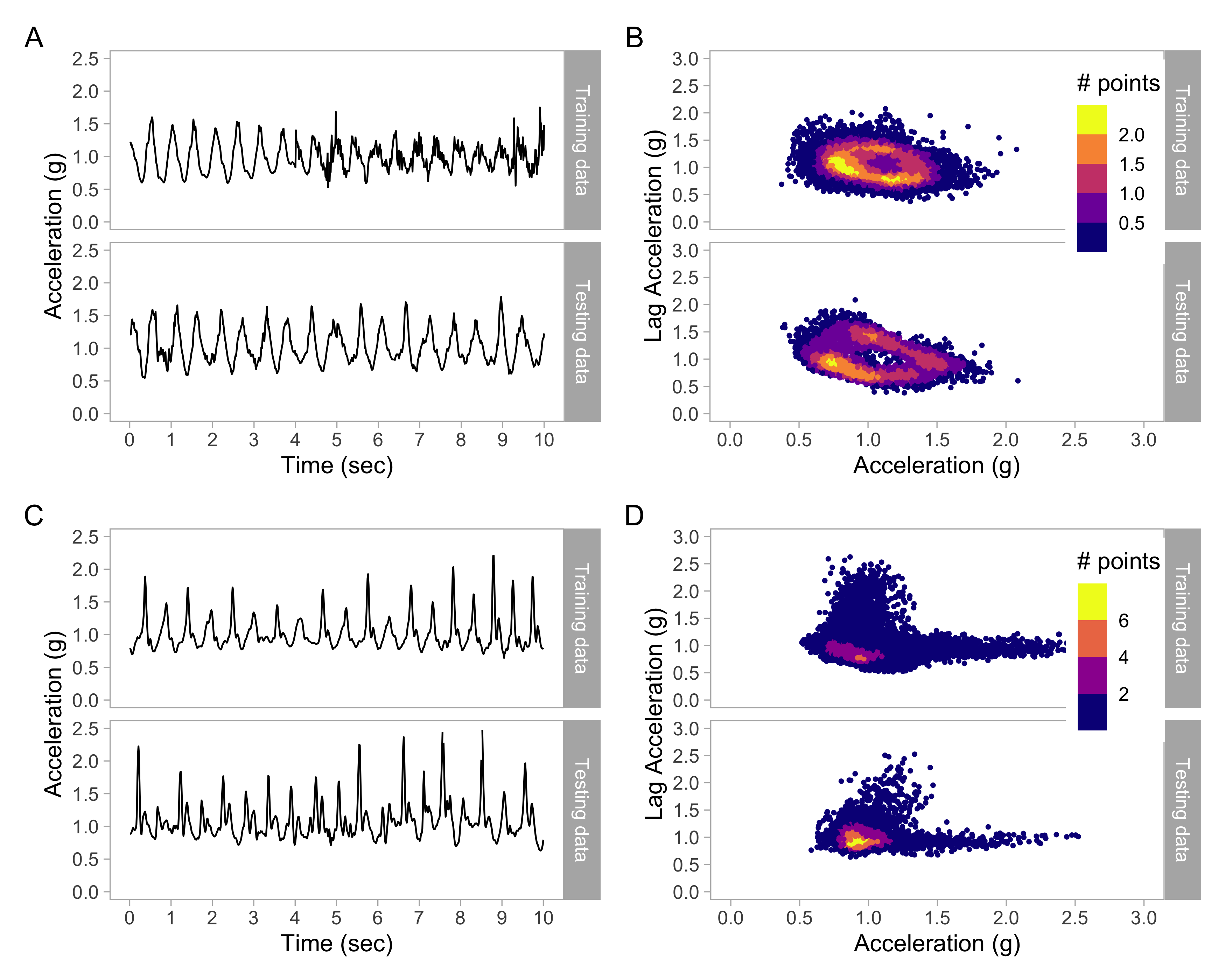}
    \caption{Raw acceleromtery from walking and associated walking fingerprints for two participants in NHANES. Panels A and C show ten seconds of walking collected on two separate days of observation for two different participants, while panels B and D show the associated walking fingerprint.}
    \label{fig:fprintdata}
\end{figure}

We use our previous algorithm for identification from wrist accelerometry data captured during walking \citep{Koffman2023, Koffman2024}. The method involves obtaining scalar summaries of the empirical joint acceleration, lag acceleration distribution and using these predictors in one versus the rest classification models; these predictors can be represented as images. Our methods achieved perfect accuracy in a dataset of $32$ individuals \citep{iu_data}, each with at least five minutes of walking and had high accuracy (98\%) in a larger dataset of $153$ individuals, each with less than one minute of walking \citep{zju_data}. In the same $153$-person dataset our method achieved moderate accuracy ($54$\%) when trained and tested on data collected at least one week apart. Figure~\ref{fig:fprintdata} demonstrates the data structure and information used for identification for two participants in the NHANES study. Panels A and C display ten seconds from the training and testing data respectively. Panels B and D display a summary of the predictors (walking fingerprints) from all training and testing data for each participant, respectively. These fingerprints represent a summary of the data that are used to predict individuals. In this paper, we will describe the process of (1) obtaining the training and testing walking data for each participant, (2) obtaining the walking fingerprint and (3) fitting models to predict the identity of the individual. 

Thus, the first challenge is identifying instances of walking for thousands of individuals with multiple days of high-resolution data. The second challenge is the size and heterogeneity of the dataset, which require the use of highly scalable methods. The third challenge is that in large dataset it is more likely that multiple individuals have similar walking patterns. To our knowledge, no existing gait-based identification approaches have solved these problems and been applied to large-scale free-living accelerometry data.

\section{Methods}\label{sec:methods}

The steps for gait-based identification in this dataset are (1) identifying walking  from up to seven full days of free living accelerometry data, (2) deriving predictors from the walking segments, and (3) fitting models to predict identity based on the predictors. 

\subsection{Segmentation of walking}
\begin{figure}[ht]
\includegraphics[width=\textwidth]{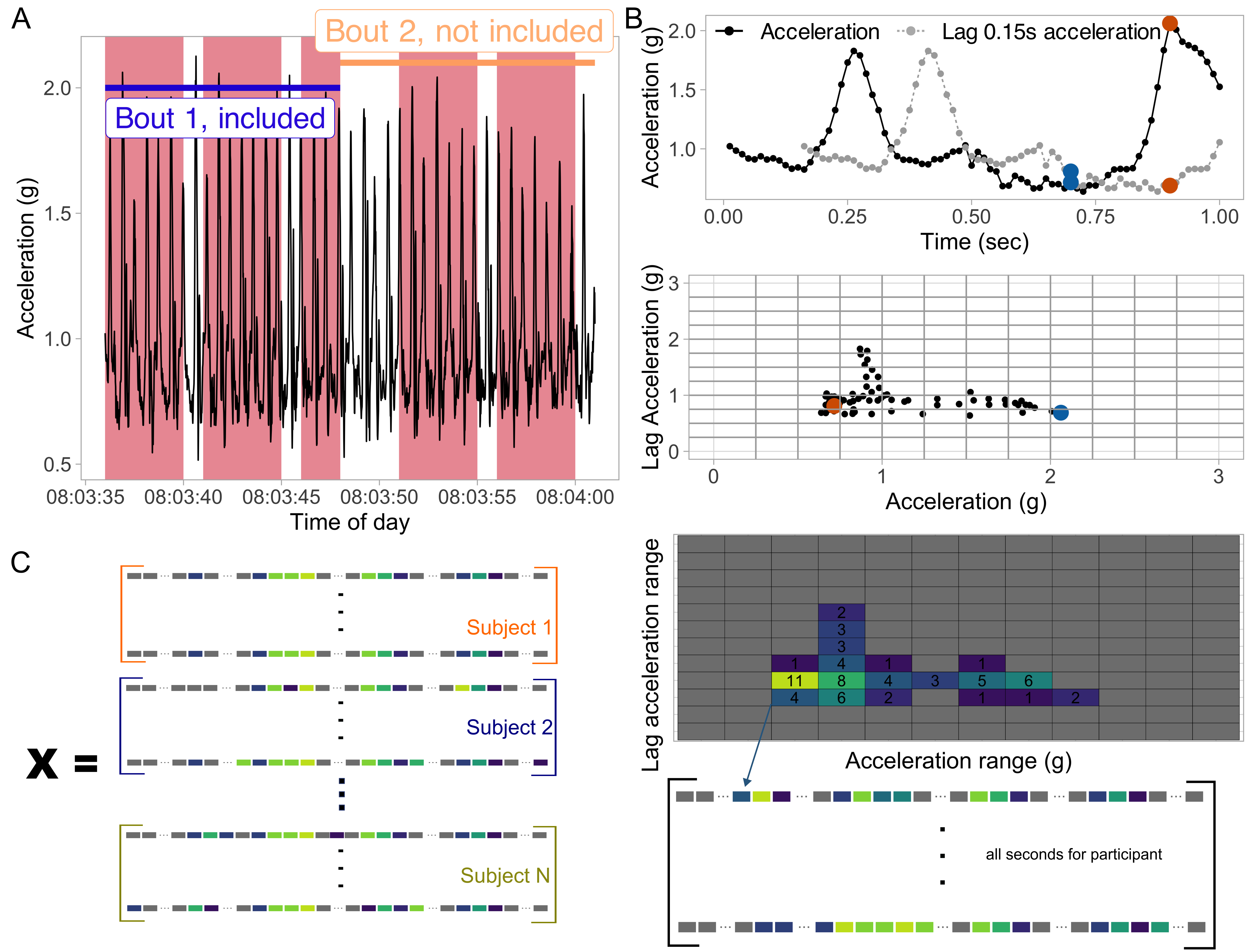}
 \caption{Deriving predictors from the raw accelerometry data for one participant. In panel A, the 30 seconds from Panel C in Figure~\ref{fig:rawacc} are shown again. This segment is comprised of two bouts. Bout 1 is 12 seconds long and has two non-consecutive seconds without walking, while bout 2 is 13 seconds and has two consecutive seconds without walking. Bout 1 is included in the analysis while bout 2 is not. In panel B, grid cell predictors are extracted for the first second of bout 1 for one lag. In panel C, the process is repeated for all participants, seconds, and lags to demonstrate how the predictor matrix is constructed.}
\label{fig:explain}
\end{figure}

The first step is to identify walking segments from over seven days of free-living accelerometry data. The majority of time for participants is \textit{not} spent walking. We used the ADEPT algorithm \citep{adept}, which is a pattern matching-based approach for stride segmentation from raw accelerometry data \citep{adept_free, stepseval}. ADEPT was applied to the raw accelerometry data for all participants to obtain an estimated number of steps per second \citep{nhanes_steps}.  Walking bouts were defined as segments of at least ten seconds in length where at least ten seconds have nonzero steps and the gap between seconds with nonzero steps is no more than one second. Individuals with less than three minutes of walking bouts over the entire observation period were excluded from analysis ($n=2{,}803; 17\%$). We chose three minutes as the minimum amount of walking time required for inclusion because our results in other datasets had shown that we could achieve high identification accuracy using three minutes of data \citep{Koffman2024} per person, and we wanted to include as many individuals in the analysis as possible. The seconds from walking bouts with nonzero steps were used for walking identification. While ADEPT provides stride segmentation (the timestamps for the beginning and end of each step), we did not use these in our methods. In fact, our methods do not require perfect walking segmentation and perform well with the estimated walking bouts by ADEPT. Panel A of Figure~\ref{fig:explain} demonstrates the extraction of walking bouts from a partial day of accelerometry data from one participant. 

\subsection{Derivation of predictors}
Predictors were obtained in a similar manner as in \citep{Koffman2024}. We briefly summarize the procedure here. Let $\mathbf{v_{ij}(s)}$ represent the acceleration in $g$ units for participant $i$ at second $j$, where $s$ denotes the samples within a second. Since there are 80 samples per second, $s \in \{1, \dots, 80\}$. Each participant has a different total number of seconds $J_i$, but for training, we set $J_i = 180 \hspace{.05in} \forall i$ (see Section~\ref{subsec:partitions}). 

For each second we compute the grid cell predictors for three lags. The grid cell predictors can be thought of as a joint histogram of the acceleration $v_{ij}(s)$ and its lag $v_{ij}(s-u)$ for three lags: $u \in \{12, 24, 36\}$. The bins for this histogram are from $0g$ to $3g$ by $0.25g$ increments of both acceleration and lag acceleration. A lag of $12$ samples at this sample rate is equivalent to a lag of $0.15$ seconds ($12$ samples/$80$ samples per second = $0.15$ seconds); thus the lags of $12$, $24$, and $36$ correspond to lags of $0.15$, $0.30$, and $0.45$ seconds used in our previous paper. 

The process of constructing the joint histogram can also be represented as constructing an image from the acceleration and lag acceleration and is shown for one second and one lag in panel B of Figure~\ref{fig:explain}. The top plot in panel B demonstrates the acceleration and lag acceleration vectors for one second and for a lag of $u=12$. The second plot in panel B demonstrates the next step in constructing the joint histogram of acceleration and lag acceleration: plotting the acceleration, lag acceleration point pairs on the acceleration by lag acceleration grid. The third plot demonstrates partitioning the grid into cells, which can be thought of a the histogram bins, and counting the number of points in each cell/bin. The fourth plot shows how these number of points in each bin become scalar predictors. Since there are $(3/0.25)\cdot(3/0.25) = 12\cdot12 = 144$ bins for each lag and $3$ lags, there are $3\cdot144 = 432$ predictors for each second and each participant.  Panel C of Figure~\ref{fig:explain} schematically demonstrates how the process of obtaining the scalar predictors is repeated for all participants and seconds. The method for identifying walking and creating grid cells is available in an R package \texttt{accelPrint} at \url{https://github.com/lilykoff/accelPrint}. 

\subsection{Partitioning of training and testing data}\label{subsec:partitions}
We consider two training/testing paradigms: random and temporal. For both, only three total minutes of data are used for each participant, and for both we employ a $75:25$ train:test split. In the random paradigm, $180$ seconds ($3$ minutes) are randomly sampled from all valid seconds (the seconds that comprise the walking bouts). Of these, $75$\% ($135$ seconds; $2$ minutes $15$ seconds) are used for training the models and the remaining $25$\% ($45$ seconds) are held out for testing the models. In the temporal paradigm, training and testing data are taken from two separate days. Training data are taken from the first day with at least $135$ seconds of walking, and testing data is taken from a later day with at least $45$ seconds of walking. The training day is always the first day with $135$ seconds of walking, while the testing day is randomly chosen from all following days with at least $45$ seconds of walking. If participants do not have a day with at least $135$ seconds of walking, and a later day with $45$ seconds of walking, they are not included in this paradigm. All individuals included in the temporal paradigm are also included in the random paradigm. There are $n = 2{,}597$ individuals included in the random paradigm but not the temporal paradigm. The temporal paradigm is meant to resemble a situation where models are trained on several days of data, and then new data is observed later in time, and we are interested in predicting the identity of the individuals in the new data. This situation may be more realistic for real-world deployment than the random paradigm, were training and testing data are mixed throughout an observation period.

\subsection{Model Fitting}
We fit classification models on the grid cell predictors derived from the training data. We evaluated performance by obtaining predictions on the grid cell predictors derived from the test data. In previous work, we found that one versus the rest classification performed well and that multinomial models were not computationally feasible. Considering the much larger sample size in this setting, we again used one versus the rest classification.  We investigated the performance of several different model types, starting on smaller subgroups of the entire sample and fitting the best-performing models from the subgroups on the entire dataset. Furthermore, we investigated strategies to improve performance, including oversampling, weighting, two-stage models, and increasing the length of the training and testing data (see Section~\ref{subsubsec:sensitivity}).

\subsubsection{Model types}\label{subsubsec:modeltypes}
In the one vs. rest classification paradigm, a separate model is fit for each participant. Six models are considered: (1) multivariable logistic regression, (2) lasso regression \citep{lasso}, (3) random forest classifier \citep{Breiman2001}, (4) gradient-boosted decision trees \citep{gradboost} using \texttt{xgboost} \citep{xgb}, (5) linear scalar-on-function regression \citep{sofr}, and (6) non-linear scalar-on-function regression \citep{sofr}. For all models except scalar on function regression, variable screening is used prior to fitting the models to remove predictors with low variance and few unique values across participants. Predictors with fewer than $10$\%  unique values across participants and with a ratio of the frequency of the most common value to the frequency of the second-most-common value greater than $95:5$ were removed; this step was implemented using the \texttt{step\_nzv} function in the \texttt{tidymodels} R package \citep{tidymodels}. As a sensitivity analysis, we investigated using principal components instead of variable screening for data reduction. For the lasso, random forest and boosted tree models, tuning parameters were selected through a parameter grid search using five-fold cross validation in the training data. The cross validation was stratified by individuals, to ensure that the proportion of data from each participant was preserved in the cross validation folds.  The parameter set with the best cross-validated area under the ROC curve (AUC) in the training data was used to fit the final model. For the linear scalar-on-function regression, the functional predictor is the 432-dimensional vector of grid cell predictors $[x_{ij1}, x_{ij2}, \dots, x_{ij432}]$ and the outcome $Y_{ij}$ is $1$ if second $(i,j)$ came from participant $i$, and $0$ otherwise. In particular, we have: 

$$g(\mu_{ij}) = \beta_0 + \int_S \beta_1(s)X_{ij}(s)ds\;, $$

\noindent where $g(\cdot)$ is the logit link function, $\mu_{ij} = \mathbb{E}[Y_{ij} | X_{ij}(s)]$,  $s = 1, \dots, 432$, $\beta_0$ is an intercept, $\beta_1(s)$ is the functional coefficient. 
In the non-linear scalar-on-function regression scenario, we instead have:

$$g(\mu_{ij}) = \beta_0 + \int_S F\{s,X_{ij}(s)ds\}\;,$$

\noindent where $F(\cdot,\cdot)$ is a smooth bivariate function. The scalar-on-function models are fit using the \texttt{pfr} function in \texttt{refund} \texttt{R} package \citep{refund}.

\subsubsection{Sample size}
This data presents two challenges compared with our previous work: lack of walking labels, and larger sample size. To investigate how both of these factors impact the model performance, we fit models on different sized subgroups of the entire population and compare results as subgroup size increases. The performance of the models in the smaller subgroups provides information about how the lack of walking labels impacts results, while the change in performance as subgroup size increases provides information about the impact of sample size. Specifically, all models were fit first on randomly selected, mutually exclusive subgroups of size $n \in \{100, 500\}$. Accuracy was calculated within each subgroup and metrics were averaged over all subgroups to evaluate model performance. The best models were then fit on subgroups of size $n \in \{1000, 2500, 5000, 10000, N = 13367\}$ for the random paradigm and  $n \in \{1000, 2500, 5000, N = 10770\}$ for the temporal paradigm and prediction performance metrics were calculated.

\subsubsection{Model evaluation and model improvements}\label{subsubsec:sensitivity} 
Within each subgroup models were evaluated by rank 1 (percent of participants correctly identified), rank 5 (percent of participants within top 5 predicted participants), rank 1\% and rank 5\% (percent of participants correctly placed in the top $1$\% and $5$\% of predictions). For example, if there are 1000 participants in the sample, the rank 1\% accuracy is the number of participants within the top 10 predictions. The metrics were averaged over subgroups to get one metric for each method and setting. 

We explored several extensions to improve our models including oversampling, weighting, and two-stage modeling. By construction, our data is class imbalanced in the one versus the rest classification models. One potential way to improve model performance is to resample with replacement in the training data for the participant the model is predicting \citep{oversampling}. This procedure increases the percent of data from the predicted participant to between $10$\% and $90$\% of the training data. Without oversampling, in the models fit on the entire population, the data from the predicted participant is less than $0.01$\% of the data $(1/13367 \cdot 100 = 0.007; 1/10770 \cdot 100 = 0.009)$. Another way to address class imbalance is weighting \citep{weighting}: we fit models where weights are constructed such that cases (the predicted participant) and controls (the not predicted participants) have equal weights. For example, if there are $n=100$ participants and each participant has $135$ observations in the dataset, then an observation from the predicted participant of interest is assigned a weight of $1/135$ and observations from the other participants are assigned a weight of $1/(135 * 99)$. 

We also used a two-stage model, where in the first step classification is performed for the entire sample. For each individual, the top $1$\% of predicted participants from the first model are selected, and a second one versus the rest classification process is carried out on just these 1\%. Final predictions are  obtained from the second model. 

All of our code for identifying walking, deriving the predictors, and fitting the models is available on Github at \url{https://github.com/lilykoff/nhanes_fingerprinting}.

\subsection{Secondary analysis with different walking identification algorithm}
As a secondary analysis, we used walking bouts derived from a different, faster, and less specific algorithm, \texttt{stepcount} \citep{stepcount}. The goal was to investigate the relative importance of specificity versus available walking time. We conducted two analyses. First, we fit logistic regression models using predictors derived from \texttt{stepcount}-identified walking bouts, using three minutes of data per person and the same individuals included in the ADEPT analysis (though not necessarily the same walking bouts), for sample sizes $n = 100$ (random and temporal paradigms) and $n = 13{,}367$ (random paradigm), and $n = 10{,}770$ (temporal paradigm). Second, we fit logistic regression models using predictors derived from \texttt{stepcount}-identified walking bouts, using 30 minutes of data per person. We performed this analysis in three different sample sizes for the random and temporal paradigms: $n = 100$, $n = \text{all individuals with 30 minutes of ADEPT-identified walking}$, and $n = \text{all individuals with 30 minutes of \texttt{stepcount}-identified walking}$. In the first two scenarios, this choice of sample size allowed us to directly compare the performance of \texttt{stepcount} and ADEPT in the same sample size.

We hypothesized that models using \texttt{stepcount}-identified walking would perform worse than the analogous models using ADEPT-derived walking bouts due to the lower specificity of \texttt{stepcount}. However, we were uncertain whether using 30 minutes of \texttt{stepcount}-derived data could offset this loss of specificity through increased sample size.

\subsection{Quantifying the impact of walking label uncertainty}
We do not know when NHANES participants are truly walking due to the free-living nature of the data. As a result, there is uncertainty in the walking labels. To quantify the impact of label uncertainty on identification accuracy, we took a simulation-extrapolation (SIMEX)-inspired approach \cite{simex} by degrading the quality of our data and comparing performance in identification accuracy. We compare four different sets of predictors generated from the following data for each participant: 
(1) random labels: 180 randomly sampled seconds of data; (2) mixed labels: 90 seconds of the randomly sampled data and 90 seconds of ADEPT-identified walking; (3) \texttt{stepcount} labels: 180 seconds of \texttt{stepcount}-identified walking; (4) ADEPT labels: 
180 seconds of ADEPT-identified walking. Using these data sets, we performed weighted logistic regression in the full dataset ($N = 13367$) and compared the resulting accuracies.

\subsection{Computational speedups}
One-vs-rest classification requires fitting $N$ separate models, which can be computationally burdensome. To address this burden, we investigated whether model training requires the full sample. We hypothesized that training on smaller numbers of individuals may be adequate, because each model needs to learn what one individual's walking looks like compared to typical population variation. We trained weighted logistic regression models on varying sized subsets of the data $n_{\text{train}} \in \{100, 500, 1000, 2500, 5000, N=13367\}$ and tested on the full data. We then compared run time and accuracy of these models.

\section{Results}\label{sec:results}
\subsection{Walking bouts and sample population}
Applying ADEPT to the full NHANES data took over $700$ days of computation time, though the method was run in parallel using 400 cores which took approximately three calendar months. Time could be further reduced by increasing the number of cores that run in parallel. Applying \texttt{stepcount} to the data took $50$ days of computation time without using parallel computing.
A total of $15{,}679$ individuals had at least one walking bout. The median (25th percentile, 75th percentile) number of bouts per participant was $64$ ($24, 141$). The median (25th percentile, 7th percentile) total time walking per participant was $17$ ($5.9$, $42$) minutes. The median (25th percentile, 75th percentile) bout length across all participants and walking bouts was $14$ ($11, 21$) seconds. These numbers correspond to the ADEPT-estimated walking, which could provide an underestimate of the total amount of walking. In our approach, we are not concerned about possible underestimation of walking time, as long as enough bouts are available to construct walking fingerprints. Visual inspection of $20$ randomly sampled bouts from ten participants confirmed that ADEPT was successful at identifying walking.

A total of $13{,}367$ individuals ($85$\%) met the inclusion criteria for the random prediction paradigm (individuals who had at least $3$ minutes of ADEPT-identified walking time). A total of $10{,}770$ individuals ($69$\%) met the inclusion criteria for the temporal prediction paradigm (individuals had at least $2$ minutes and $15$ seconds in the first days and at least $45$ seconds at a later day of ADEPT-identified walking). A total of $15{,}374$ individuals (95\%) met the criteria for the \texttt{stepcount} analysis (30 minutes of walking time with at least 22.5 minutes on one day and 7.5 minutes on a later day; these individuals are used for both the random and temporal paradigm with \texttt{stepcount}). Unweighted summaries of the included individuals in each dataset are provided in Table~\ref{tab:popsummary}. After variable screening, the median [minimum, maximum] number of variables included in the datasets used for model fitting was $77$ [69, 80]. Excluding the walking labeling step, creation of the data used in the models took approximately 10 hours of computation time. After data creation, model fitting for logistic regression on the entire dataset (between $10{,}000$ and $16{,}000$ individuals, depending on the task) took 6000 hours for models using 30 minutes of data per person and 1000 hours for models using 3 minutes of data per person. Using 400 cores and 2.5 Tb of RAM, these tasks took 120 hours and 23 hours, respectively.

\begin{table}[H]
\begin{tabular}[t]{lllll}
\toprule
 & Overall & Random & Temporal & \texttt{stepcount}\\
&$N = 16{,}170$ &  $N = 13{,}367$ &  $N=10{,}770$  & $N=15{,}374$\\
\midrule
Data release cycle &  &  & \\
\hspace{.05in}Wave G, 2011-2012 & 6,917 (43\%) & 5,899 (44\%) & 4,897 (45\%) & 6,672 (43\%)\\
\hspace{.05in}Wave H, 2013-2014 & 7,776 (48\%) & 6,336 (47\%) & 5,103 (47\%) & 7,254 (47\%)\\
\hspace{.05in}NNYFS & 1,477 (9.1\%) & 1,132 (8.5\%) & 770 (7.1\%) & 1,448 (9.4\%)\\
Sex female & 8,245 (51\%) & 6,760 (51\%) & 5,295 (49\%) & 7,808 (51\%)\\
Age & 33 (23) & 33 (22) & 35 (21) & 33 (23)\\
Race &  &  & \\
\hspace{.05in}Non-Hispanic White & 5,736 (35\%) & 4,625 (35\%) & 3,705 (34\%)& 5,425 (35\%)\\
\hspace{.05in}Non-Hispanic Black & 3,964 (25\%) & 3,176 (24\%) & 2,463 (23\%)&  3,746 (24\%)\\
\hspace{.05in}Mexican American & 2,413 (15\%) & 2,095 (16\%) & 1,699 (16\%) & 2,310 (15\%)\\
\hspace{.05in}Other Hispanic & 1,663 (10\%) & 1,422 (11\%) & 1,187 (11\%)&  1,597 (10\%)\\
\hspace{.05in}Other Race - Including Multi-Race & 2,394 (15\%) & 2,049 (15\%) & 1,716 (16\%) & 2,296 (15\%)\\
Weight (kg) & 67 (29) & 68 (27) & 71 (25) & 66 (28)\\
\hspace{.05in}\textit{Missing} & 115 & 52 & 43 & 69\\
Height (cm) & 157 (20) & 160 (18) & 162 (16) & 158 (20)\\
\hspace{.05in}\textit{Missing} & 120 & 51 & 43  & 69\\
BMI (kg/m$^2$) & 26 (8) & 26 (7) & 26 (7) &  25 (8)\\
\hspace{.05in}\textit{Missing} & 139 & 56 & 46  & 80\\
\bottomrule
\end{tabular}
\caption{Characteristics of the sample. The overall column consists of all individuals who had any accelerometer data available for analysis, while the random and temporal columns describe individuals included in each paradigm, respectively. Continuous variables presented as mean (SD), categorical variables presented as n (\%)}
\label{tab:popsummary}
\end{table}

\subsection{Model performance in subgroups of size n = 100 and n = 500}
\begin{figure}[H]
    \centering
    \includegraphics[width=\linewidth]{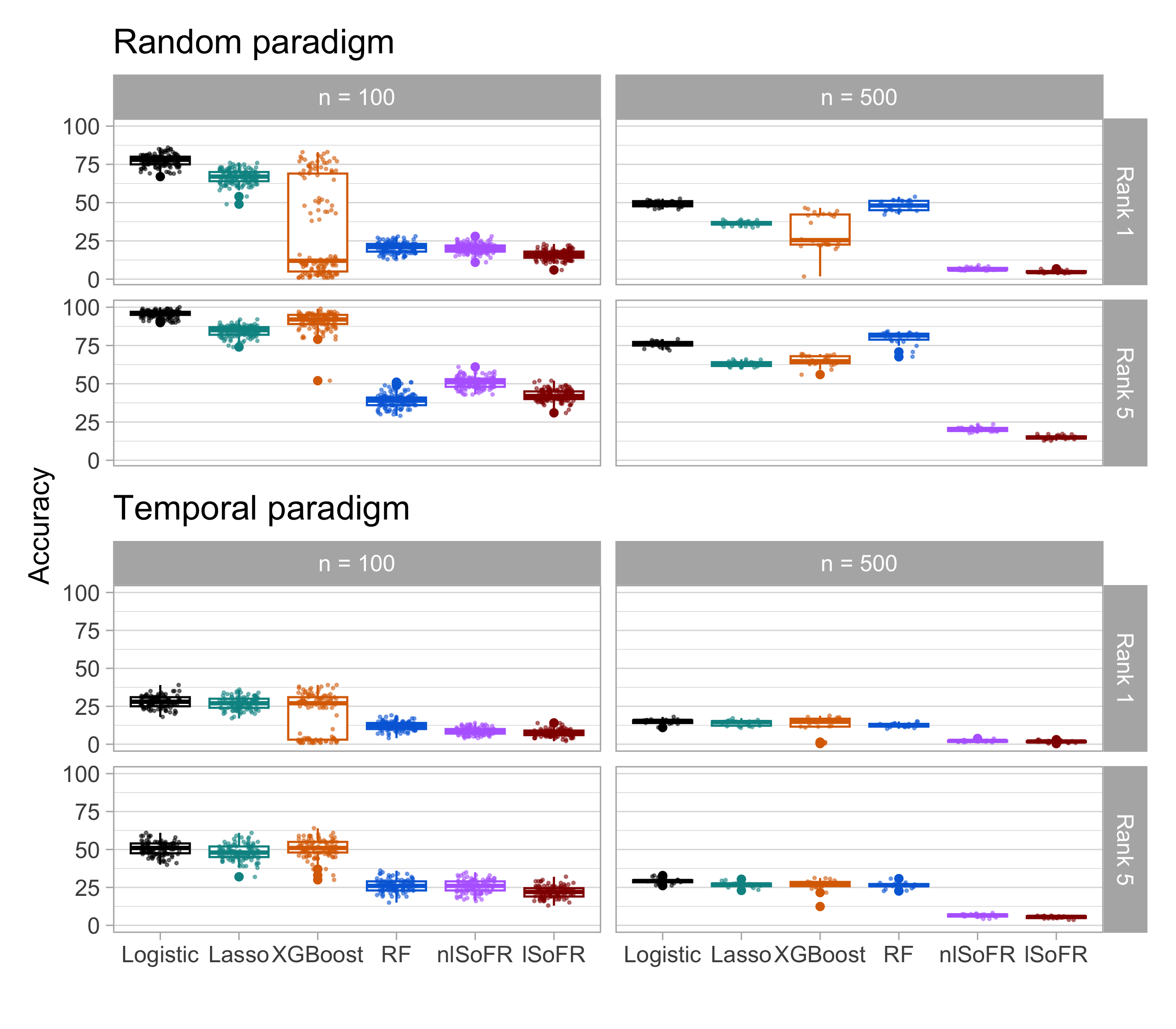}
    \caption{Rank 1  and rank 5 accuracies for each model type fit on subgroups of $n = 100$ (left panel) and $n = 500$ (right panel), for random train/test splits (top panel) and temporal train/test splits (bottom panel). Logistic regression performs best across all scenarios. RF = random forest, nlSoFR = nonlinear scalar on function regression, lSoFR = linear scalar on function regression.}
    \label{fig:acc100}
\end{figure}

All classification models: logistic regression, lasso, random forest, XGBoost, and scalar-on-function regression were run on subgroups of size $n=100$. Each participant belongs to only one subgroup: the subgroups are mutually exclusive. In the random paradigm, there were $133$ such subgroups, and the $13367-(133*100) = 67$ participants who did not fit evenly into one of the groups were excluded. In the temporal paradigm there were $107$ subgroups of size $100$. Rank 1 and rank 5 accuracy were calculated in each subgroup and then averaged over subgroups to obtain one performance metric for each model. The same process was repeated for subgroups of size $n=500$. Table~\ref{tab:acc100} presents the median, minimum, and maximum accuracies across all of the subgroups, while Figure~\ref{fig:acc100} displays each model's performance in the individual subgroups. 

For $n=100$, in both the random and temporal settings, logistic regression has the highest median rank 1 accuracies ($78$\%, $28$\%) and the highest rank 5 accuracies ($96$\%, $51$\%). Lasso has the second-highest rank 1 ($67$\%) and third-highest rank 5 ($85$\%) accuracy for the random paradigm; for the temporal paradigm, lasso is tied with XGBoost the second-best rank 1 accuracy ($27$\%) and has the second-best rank 5 accuracy ($48$\%). The minimum rank 1 accuracy for XGBoost ($1$\% in both random and temporal) is quite low, indicating poor performance in some subgroups; this is clear from the large spread of the boxplots for XGBoost in Figure~\ref{fig:acc100} as well. For all models except XGBoost, rank 1 accuracy is between $43$\% and $64$\% lower in the temporal setting than the random setting. For all models, rank 5 accuracy is between $40$\% and $70$\% lower in the temporal setting than the random setting. Note that in this context ($n=100$) rank 1 and rank 1\% accuracy are the same, as are rank 5 and rank 5\% accuracy.

For $n=500$, logistic regression again had the highest or tied for highest median rank 1 accuracy in both the random and temporal paradigms ($50$\%, $15$\%). Logistic regression had the highest rank 5 accuracy in the temporal paradigm ($29\%$), while the random forest had the highest rank 5 accuracy in the random paradigm (82\%). The scalar-on-function models performed much worse than logistic regression, lasso regression, or the machine learning models. Interestingly, the random forest performed as well or better for $n=500$ than for $n=100$ in both the random and temporal paradigms.  For all models, rank 1 and rank 5 accuracy is lower in the temporal setting than the random setting, as is rank 1\% and rank 5\% accuracy. Note that in this context ($n=500$), rank 1\% accuracy is the same as rank 5 accuracy).

\begin{table}[H]
\begin{tabular}[t]{lllllll}
\toprule
& & \multicolumn{2}{c}{Rank 1} & \multicolumn{2}{c}{Rank 5}  & Rank 5\%  \\
 & Model & $n=100$ &$n=500$ & $n=100$  &$n=500$  &$n=500$ \\
\cmidrule(lr){1-7}
\multirow{5}{*}{Random} & Logistic & \textbf{78 [67,86]} & \textbf{50 [46,53]} & \textbf{96 [90,100]} & 76 [72,79] &  95 [93,97]\\
& Lasso & 67 [49,76] & 36 [34,39] & 85 [74,92] & 63 [61,66]  & 89 [86,91]\\
 & Random Forest & 21 [13,28] & 48 [42,54] & 39 [29,51] & \textbf{82 [68,84]} &\textbf{97 [92,99]}\\
& Nonlinear SoFR & 20 [11,28] & 6.2 [5.6,9.2] & 51 [43,61] & 20 [18,24]  & 49 [44,51]\\
 & Linear SoFR & 16 [6,23] & 4.6 [3.8,6.8] & 42 [31,52] & 15 [13,17] & 40 [37,43]\\

 & XGBoost & 12 [1,83] & 26 [1.8,47] & 92 [52,99] & 65 [56,69] & 92 [89,94]\\
 \cmidrule(lr){2-7}
\multirow{5}{*}{Temporal} & Logistic & \textbf{28 [18,39]} & \textbf{15 [11,18]} & \textbf{51 [40,61]} &  \textbf{29 [26,33]} & \textbf{51 [46,54]}\\
 & Lasso & 27 [17,36] & 15 [11,17] & 48 [32,61] & 27 [23,30] &  46 [41,52]\\
 & Random Forest &  12 [4,19] & 13 [10,15] & 26 [15,36] & 26 [23,31] & 49 [47,53] \\
 & Nonlinear SoFR &  9 [4,14] & 2.2 [1.2,3.8] & 26 [16,35] &  6.6 [4.2,8.2] & 19 [15,21]\\
 & Linear SoFR & 7 [2,14] & 1.8 [0.4,3] & 22 [13,32] & 5.6 [3.6,6.4] &  16 [11,18]\\
 & XGBoost & 27 [1,39]  & 15 [0.4,19] & 51 [30,64] & 27 [12,31] & 50 [45,52]\\
\bottomrule
\end{tabular}
\caption{Rank 1, rank 5, and rank 5\% median [minimum, maximum] accuracies of models on subgroups of size $n=100$ and $n=500$. Note: rank 1\% and rank 5\% accuracies for $n=100$ are not shown because they are equivalent to rank 1 and rank 5 accuracies for $n=100$; likewise rank 1\% accuracy is equivalent to rank 5 accuracy for $n=500$ and is not shown. The best model in each category is bolded.}
\label{tab:acc100}
\end{table}

\subsection{Model performance in larger subgroups}\label{subsec:largersubgroups}

\begin{figure}[ht]
    \centering
    \includegraphics[width=\linewidth]{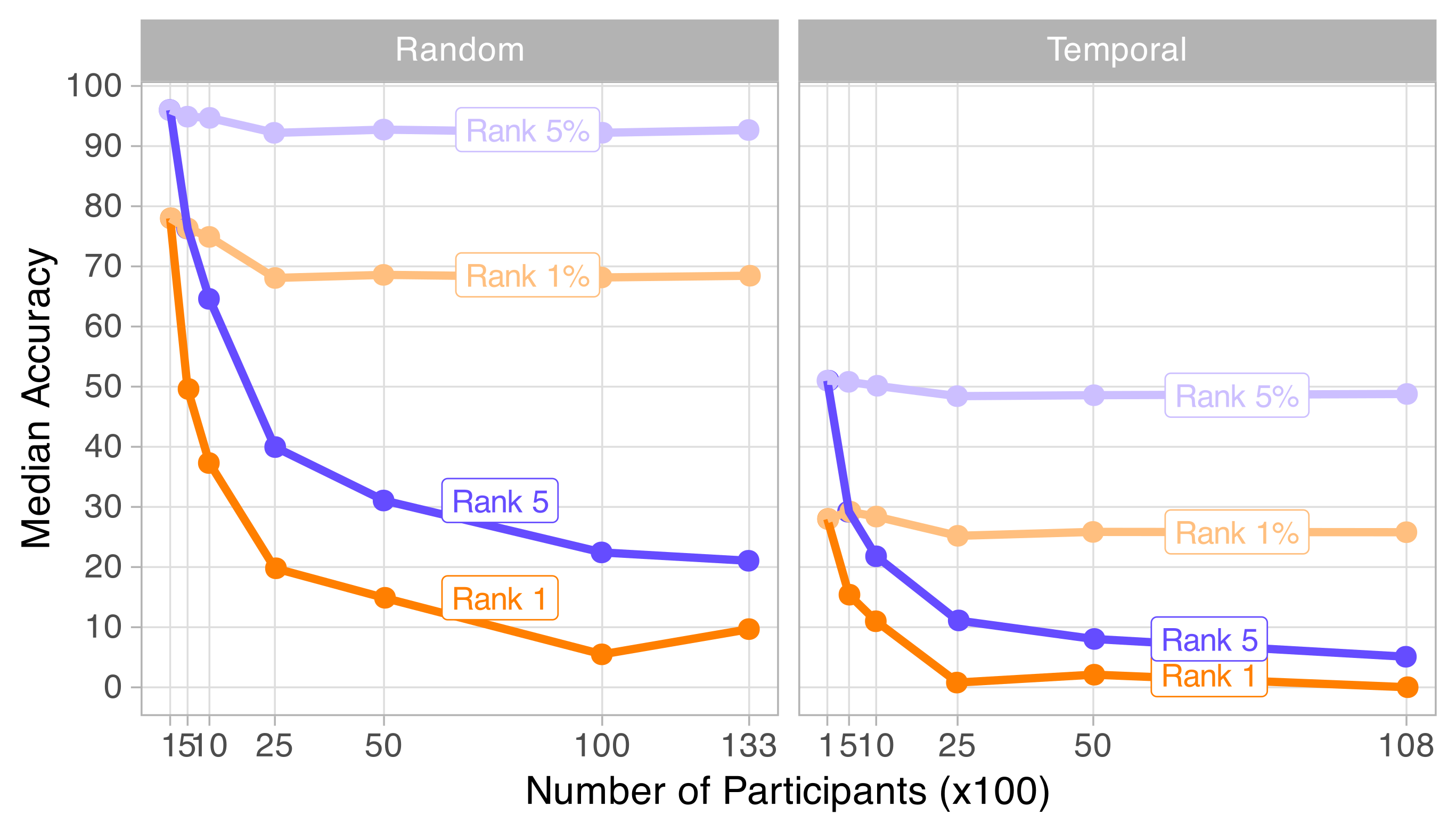}
    \caption{Rank 1, rank 5, rank 1\%, and rank 5\% accuracies for logistic regression models with varying size subgroups for random (left panel) and temporal (right panel) models.}
    \label{fig:accov}
\end{figure}

\begin{figure}[ht]
    \centering
    \includegraphics[width=\linewidth]{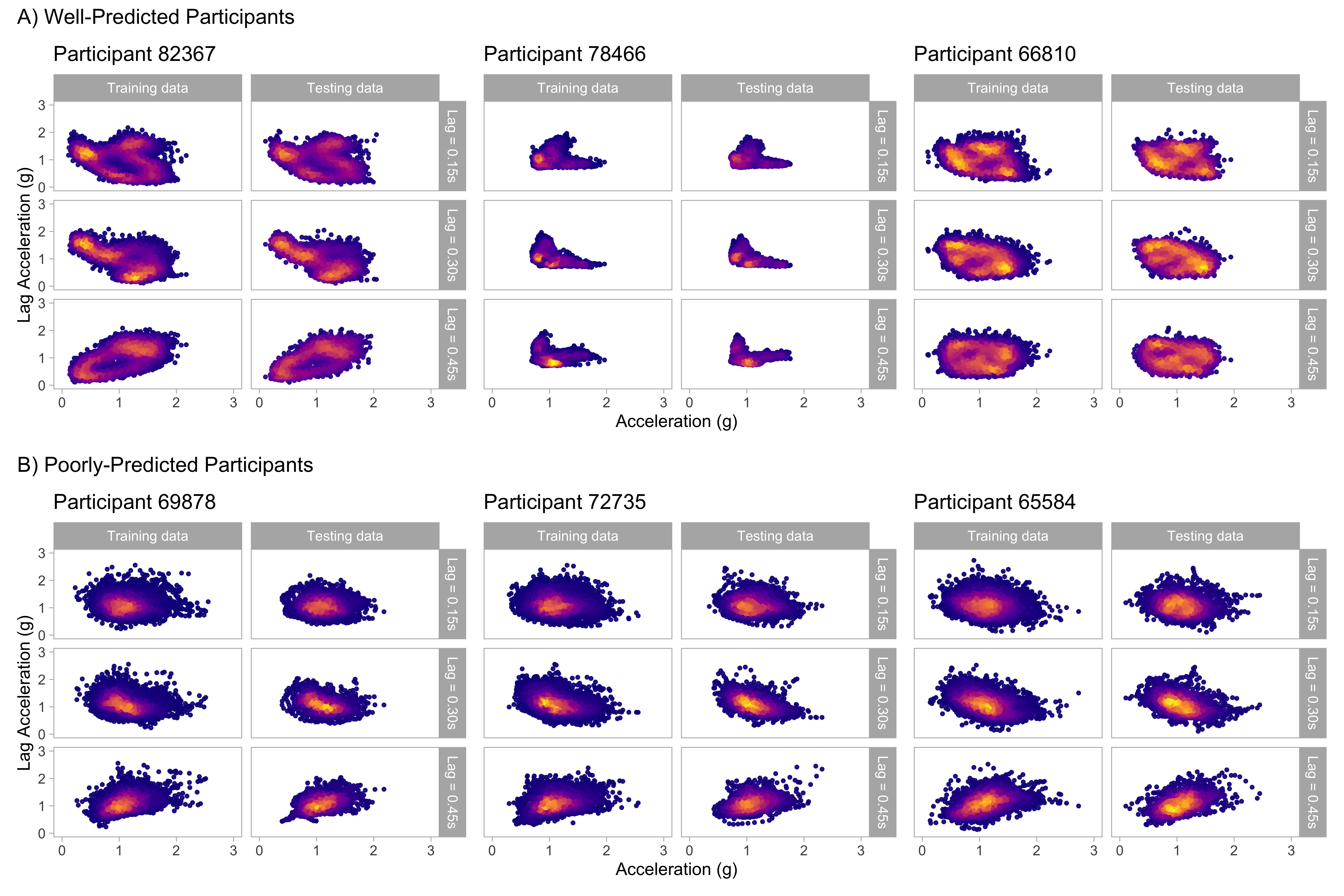}
    \caption{Summaries of the walking fingerprint in training and testing data for participants who were well-predicted in a subgroup of $n=100$ (panel A) and participants who were not correctly predicted in a subgroup of $n=100$ (panel B). Each facet is a summary of the walking fingerprint for a specific participant, lag, and training or testing data.}
    \label{fig:predictions}
\end{figure}

The random forest, XGBoost, and lasso models require hyperparameter tuning. Using five-fold cross-validation, this requires fitting each model six times for each participant and hyperparmeter combination, which is computationally expensive even for moderate sample sizes. We fit the random forest on ten subgroups of size $n = 1000$ in the random paradigm, which took over 96,000 hours of computation time. Even with parallelization on a computing cluster, these models took over two months to run (and still represent just a subset of the entire population); performance of the random forest models was similar to the logistic regression models (see Table~\ref{tab:acc_all}). While the scalar on function regression models were less computationally expensive, they had poor performance in smaller subgroups. Therefore only the logistic regression models were fit on $n > 500$.
Figure~\ref{fig:accov} shows rank 1, rank 5, rank 1\%, and rank 5\% accuracy for logistic regression models on subgroups varying from size $n=100$ to $n = 13{,}367$ (random paradigm) or $n=10{,}770$ (temporal paradigm), and Table~\ref{tab:acc_all} shows the performance at each sample size in more detail. Rank 1 and rank 5 accuracy decrease with increasing sample size: in the random setting, rank 1 accuracy decreases from $78$\% to $10$\%, in the temporal setting, it decreases from $28$\% to $0$\%. However, the rank 1\% and rank 5\% accuracies stay relatively constant: $78$\% to $68$\% in the random setting; $26$\% to $24$\% in the temporal setting for rank 1\%. Thus regardless of sample size, in the random setting, the correct participant is in the top 5\% of predictions $93$\% of the time and in the top 1\% of predictions $68$\% of the time. Figure~\ref{fig:predictions} provides some intuition for the performance of the models. In each sub-panel, the data in the left column is a summary of training data, the data in the right column is a summary of testing data. Among well-predicted participants (panel A), training and testing data look similar within individuals and different between individuals. Among poorly-predicted participants (panel B), while training and testing data are not dissimilar within individuals, training and testing data are also similar \textit{between} individuals, so models are not able to correctly identify these participants.

\begin{table}
\begin{tabular}[t]{lllllll}
\toprule
 & Sample size & Number of subgroups & Rank 1 & Rank 1\% & Rank 5 & Rank 5\% \\
\cmidrule(lr){1-7}
\multirow{8}{*}{Random} & 100 & 133 &  78 [67,86] & 78 [67,86] & 96 [90,100] & 96 [90,100]\\
 & 500 & 26 & 50 [46,53] & 76 [72,79] & 76 [72,79] & 95 [93,97]\\
 & 1000 & 13 & 37 [35,40] & 75 [72,76] & 65 [61,67] & 95 [94,96]\\
 & 1000* & 10 & 38 [41,45] & 84 [81, 87] & 73 [69, 77] & 98 [97, 98]\\
 & 2500 & 5 & 20 [4.3,20] & 68 [68,69] & 40 [37,43] & 92 [92,93]\\
 & 5000 & 2 & 15 [14,16] & 69 [68,69] & 31 [31,31] & 93 [93,93]\\
 & 10000 & 1 & 5.5  & 68  & 22 & 92 \\
& 13367 & 1 &  9.7  & 68  & 21  & 93 \\
 \cmidrule(lr){2-7}
\multirow{6}{*}{Temporal} &  100 & 107 & 28 [18,39] & 28 [18,39] & 51 [40,61] & 51 [40,61]\\
 & 500 & 21 & 15 [11,18] & 29 [26,33] & 29 [26,33] & 51 [46,54]\\
 & 1000 & 10 & 11 [1,13] & 28 [27,30] & 22 [20,23] & 50 [49,52]\\
 & 2500 &5 &  0.8 [0.08,5.4] & 25 [25,27] & 11 [10,12] & 48 [48,49]\\
 & 5000 & 2 & 2.1 [0.02,4.2] & 26 [25,26] & 8 [7.2,8.9] & 49 [48,49]\\
 & 10770 & 1 & 0.028 & 26  & 5.1 & 49 \\
\bottomrule
\end{tabular}
\caption{Rank 1, rank 1\%, rank 5, rank 5\% median [minimum, maximum] accuracies of logistic regression models for all sample sizes. If only one subgroup of models is fit, the minimum and maximum are not shown. The asterisk in the fourth row denotes results from the random forest fit on ten subgroups of size 1000, for comparison to the logistic regression model.}
\label{tab:acc_all}
\end{table}

\subsection{Model improvements}
Table~\ref{tab:acc_large} summarizes the various model improvements tested in the random and temporal paradigms on the full dataset. The top row describes the performance without any model improvements. In the random paradigm, oversampling at 10\% and weighting perform best, improving rank 1 accuracy from 9.7\% to 41\% and 34\%, respectively. ``Weighted with PCA'' in the random paradigm refers to a weighted logistic regression model using scores from the principal components that explained 90\% of the variation in  the test data as predictors; it exhibits much lower accuracy than the other models. In the temporal paradigm, the two-stage model works best, improving rank 1 accuracy to 5.2\% from 0.03\%. 
\begin{table}[H]
\begin{tabular}[t]{llllll}
\toprule
 & Model & Rank 1 & Rank 1\% & Rank 5 & Rank 5\%\\
\cmidrule(lr){1-6}
\multirow{4}{*}{{\shortstack{Random\\$n=13{,}367$}}} & Logistic &  9.7 & 68 & 21 & 93\\
&  Oversampled at 10\% &   \textbf{41}   & 68 &     \textbf{95}    &   99 \\
& Weighted & 34 & \textbf{96} & 61 & \textbf{100}\\
& Weighted with PCA & 1.2 & 28 & 4.7 & 50 \\
& Two-stage &  20 & 68& 37 & 93 \\

 \cmidrule(lr){2-6}
 \multirow{4}{*}{{\shortstack{Temporal\\$n=10{,}770$}}} 
 & Logistic &  0.028 & 26 & 5.1 & 49\\
 & Oversampled at 10\% &  4.3 & \textbf{32} & \textbf{10} & \textbf{52}\\
 &Weighted &  1.8 & 23 & 5.1 & 45\\
 & Two-stage &    \textbf{5.2} &26 & 10   & 49 \\
\bottomrule
\end{tabular}
\caption{Rank 1, rank 1\%, rank 5, rank 5\% accuracies of different model types on the entire population for each model. The best model in each category is bolded.}
\label{tab:acc_large}
\end{table}

\subsection{Different walking identification algorithm}

Table~\ref{tab:scvadept} compares results from models fit on predictors derived from walking bouts identified by ADEPT versus \texttt{stepcount}, for various sample sizes and minutes of data: the top half displays results using 3 minutes of data per person, the bottom half displays results using 30 minutes of data per person. Comparison across rows indicates that, for the same sample size and minutes of data, ADEPT outperforms \texttt{stepcount} (e.g.: 78\% vs. 23\% rank 1 accuracy for $n=100$, random paradigm with 3 minutes of data, 97\% vs. 50\% rank 1 accuracy for $n=100$, random paradigm with 30 minutes of data. This pattern holds for all paradigms, sample sizes, and minutes of data. Comparing within columns indicates that, holding method, paradigm, and sample size constant, increasing minutes used improves performance (e.g. 78\% vs. 97\% rank 1 accuracy, random paradigm, $n=100$ for ADEPT; 23\% vs. 50\% rank 1 accuracy, random paradigm, $n=100$ for \texttt{stepcount}. 
The top half of Table~\ref{tab:scvadept} compares results from models fit on predictors derived from walking bouts identified by ADEPT versus \texttt{stepcount}, using three minutes of data per person. The individuals included in the models are the same but the walking bouts are not. Results from ADEPT are better for subgroups of size $n = 100$ in both the random paradigm (median rank 1 accuracy = $78$\% for ADEPT, $50$\% for \texttt{stepcount}) and the temporal paradigm (median rank 1 accuracy = $26$\% for ADEPT, $17$\% for \texttt{stepcount}). This pattern holds for the larger sample sizes as well: \texttt{stepcount} has lower rank 1 and rank 5 accuracy compared to ADEPT in the random and temporal paradigms for both logistic regression and weighted logistic regression.

Finally, comparing ADEPT with 3 minutes of data to \texttt{stepcount} with 30 minutes of data in similar sample sizes (row 3 vs. row 9 and row 6 vs. row 12 of Table~\ref{tab:scvadept}), indicates that \texttt{stepcount} with 30 minutes of data outperforms ADEPT with 3 minutes of data (34\% vs. 65\% rank 1 accuracy, random paradigm, and 1.8\% vs. 3.5\% rank 1 accuracy, temporal paradigm).

\begin{table}[H]
\begin{tabular}[t]{lllrllll}
\toprule
 & &\multirow{2}{*}{Method}  & \multirow{2}{*}{Sample size} & \multicolumn{2}{c}{Rank 1} & \multicolumn{2}{c}{Rank 5} \\
& & & & ADEPT & \texttt{stepcount} & ADEPT & \texttt{stepcount} \\
\midrule
\multirow{6}{*}{3 min} & 
\multirow{3}{*}{Random} & Logistic & 100 & 78 [67,86] & 23  [15,33] & 96 [90,100] & 52 [39,64]\\
& &Logistic & 13367 & 9.7  & 1.3  & 21  & 3.7 \\
&& Logistic; weighted & 13367 & 34 & 1.2 & 61  & 4.4 \\
\cmidrule{2-8}
&\multirow{3}{*}{Temporal} & Logistic & 100 & 26 [2,38] & 5 [1,11] & 50 [38,62] & 19 [9,29]\\
&& Logistic & 10770 & 0.028 & 0.028  & 5.1 & 0.33 \\
&& Logistic; weighted & 10770 & 1.8  & 0.093  & 5.1  & 0.36 \\
\midrule \midrule 
\multirow{6}{*}{30 min} & \multirow{3}{*}{Random} & Logistic & 100 & 97 [92,100] & 50 [37,61] & 100 [100,100] & 88 [79,95]\\
&& Logistic & 5302 & 32 & 7.7  & 57 & 20 \\
&& Logistic; weighted & 15374 & -- & 65 & --  & 86 \\
\cmidrule{2-8}
&\multirow{3}{*}{Temporal} & Logistic & 100  & 59 [46,69] & 17 [10,25] & 79 [74,87] & 43 [33,54]\\
&& Logistic & 1541 & 27 & 2.9 & 45& 7.2 \\
&& Logistic; weighted & 15374 & -- & 3.5 & --  & 8.1 \\
\bottomrule
\end{tabular}
\caption{Comparison of accuracies for regression models fit on predictors from ADEPT vs. \texttt{stepcount}. Accuracies are presented as median [minimum, maximum] if there are multiple subgroups.}
\label{tab:scvadept}
\end{table}

\subsection{Quantifying the impact of walking label uncertainty}
Table~\ref{tab:scvadept} presents the results of the analysis to investigate label uncertainty. The random labels dataset achieves less than 1\% rank 1 accuracy, while the mixed labels dataset achieves 9\% rank 1 accuracy. The random labels dataset performs worse than both the \texttt{stepcount} labels dataset (1\% rank 1 accuracy) and the ADEPT labels dataset (34\% rank 1 accuracy), while the mixed label dataset performs better than \texttt{stepcount} but worse than ADEPT.

\begin{table}[H]
\begin{tabular}{lllll}
\toprule 
Model     & Rank 1 & Rank 1\% & Rank 5 & Rank 5\% \\
\midrule 
Random labels & 0.13   & 12       & 0.90   & 34       \\
Mixed labels    & 9      & 70       & 22     & 92       \\
\texttt{stepcount} labels & 1      & 31       & 4      & 63       \\
ADEPT labels    & 34     & 96       & 61     & 100     \\
\bottomrule
\end{tabular}
\caption{Comparison of accuracies for  weighted logistic regression models fit on four different sets of predictors comprised of three minutes of data from each participant ($N=13367)$. Random labels are randomly selected seconds, mixed labels are 50\% random labels and 50\% ADEPT-identified walking, \texttt{stepcount} labels are \texttt{stepcount}-identified walking, and ADEPT labels are ADEPT-identified walking}
\end{table}

\subsection{Computational speedups}
\begin{figure}[ht]
    \centering
    \includegraphics[width=\linewidth]{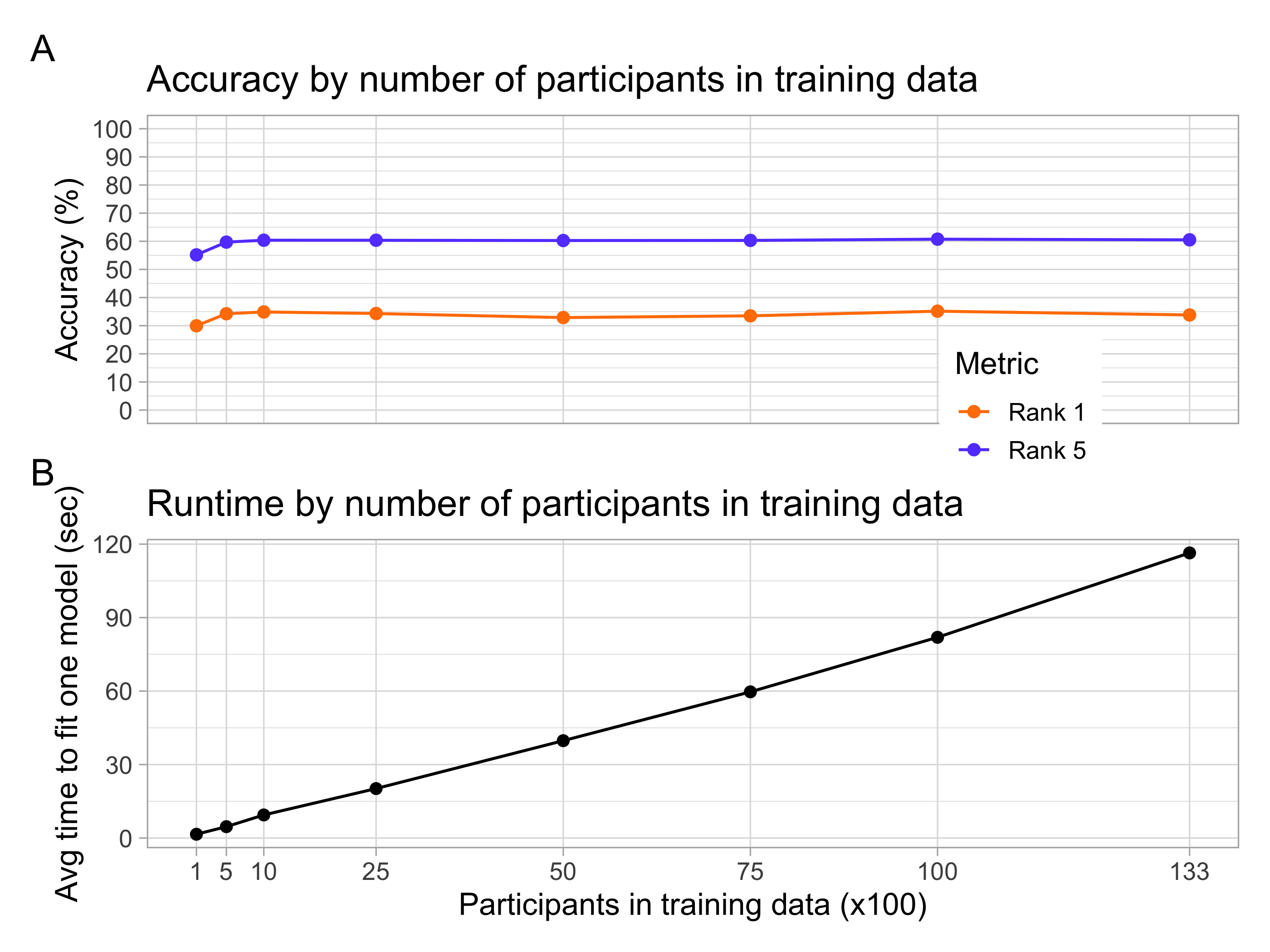}
    \caption{Accuracy (A) and runtime (B) of weighted logistic regression models trained on datasets with sample sizes varying from 100 to 13367 and tested on the full dataset.}
    \label{fig:subsets}
\end{figure}

Figure~\ref{fig:subsets} displays accuracy and runtime for weighted logistic regression models fit with different numbers of individuals in the training data. Runtime increases linearly with number of participants in the training data: runtime is 1.6 seconds per model when $n_{\text{train}} = 100$ compared to 116 seconds per model when $n_{\text{train}} = 13367$. Improvements in accuracy plateau above $n_{\text{train}} = 500$: rank 1 accuracy remains between 32.5\% and 35.2\%, while rank 5 remains between 59.7\% and 60.8\%.

\section{Discussion}
While the accuracy may need to be improved for use of true identification, our approach compares favorably to other large-scale biometric identification efforts using free-living data. For instance, speaker identification using convolutional neural networks on YouTube interviews (VoxCeleb, $n\approx 1{,}200$) achieved approximately $80$\% rank 1 accuracy \citep{voxceleb}; our method achieved roughly $75$\% rank 1 accuracy in a comparable task with $n=1{,}000$. Large-scale facial recognition on unconstrained images (MegaFace) reports around $75$\% accuracy with one million distractors \citep{megaface}. Biometric identification from wearable ECG in daily life has achieved approximately $90$\% accuracy, though in much smaller samples ($n=20$) \citep{ecg_wild}. By contrast, the number of individuals expected to be correctly identified by random guessing in our full sample ($n=13{,}367$) is at most $3$ or $4$ \citep{poisson_fprint}, underscoring the strong signal captured by our approach.

Our approach requires first segmenting walking from the data using ADEPT, then calculating grid-cell predictors (the joint histogram of acceleration and lag acceleration) from a sample of the walking data from each participant. After variable screening, one versus rest classification models were then fit using the grid cell predictors on varying size subgroups of the data. Two train/test paradigms were employed: random subsampling, and temporal sampling, where models are trained on data from one day and predicted on another day. Interestingly, we found that using principal components for data reduction led to significantly worse performance than removing predictors with low variance. We hypothesize this is because PCA finds features that explain population-level variance and may obscure person-specific gait features that distinguish between individuals.

Across all subgroups and training/testing paradigms, multivariable logistic regression outperformed regularized (lasso) logistic regression, machine learning models (random forest and XGBoost) and scalar on function regression models. The correct participant was in the top 1\% of predictions $96$\% of the time and in the top 5\% of predictions $100$\% of the time for the weighted model fit on all of the data (random), and $32$\% of the time and $52$\% of the time for the oversampled model on the temporal data. Of all the modeling approaches, we recommend weighted logistic regression for the full dataset because it balances ease of implementation, computational efficiency, and performance across train/test paradigms.

In subgroups of size $n=100$, we achieved $78$\% accuracy (random paradigm) and $28$\% accuracy (temporal paradigm), which is somewhat lower than the accuracy achieved in a similar sized dataset with ground truth walking labels ($n=153$, rank 1 accuracy of $93$\% and $41$\%  for scenarios corresponding to the random and temporal paradigms, respectively). By increasing the amount of data used for training and testing to $30$ minutes from $3$ minutes, rank 1 accuracy is $97$\% (random) and $59$\% (temporal). The results in the $n=100$ subgroup indicate that the lack of walking labels does impact results, but person identification is still very feasible using ADEPT walking bouts, especially if a larger number of bouts is used. Performance decreased substantially with increasing number of individuals in the model, indicating that the chief challenge in this data comes from the larger sample, not the lack of walking labels. However, when accuracy was normalized for sample size, performance stayed relatively constant. Furthermore, using 30 minutes of data derived from a different walking identification algorithm, \texttt{stepcount}, led to higher accuracy (65\% rank 1) in a large sample size ($n=15{,}374$, indicating that perhaps a method that balances the specificity of ADEPT with the availability of walking provided by \texttt{stepcount}, trained using at least 30 minutes of data, is the best approach. Model improvement techniques including oversampling and weighting led to large improvements in accuracy for the random paradigm and smaller improvements in accuracy for the temporal paradigm. This indicates that poor performance in the random paradigm is be partially due to class imbalance, while poor performance in the temporal paradigm is more attributable to the fact that walking is inherently different on different days.  Walking can differ temporally due to differences in mood, fatigue, daily patterns, or accelerometer device placement. Previous research has found that gait speed and surface type can have large influence on gait recognition \citep{gaitspeed}. 

Analyses to quantify the impact of label uncertainty on identification accuracy illustrated important information. First, label accuracy matters: we see substantial decrease in model performance from ADEPT (accurate labels) to random labels (inaccurate labels). Second, the ADEPT labels are fairly robust to moderate contamination: by mixing an equal proportion of ADEPT labels with random labels, we still achieve 9\% rank 1 accuracy and 22\% rank 5 accuracy vs. 34\% and 61\% with perfect labels, respectively. As expected, very low accuracy is achieved with the random labels dataset, which confirms that the ADEPT-based data is capturing walking patterns and not other movement patterns throughout the day, and that walking data in this context is crucial for identification.  Future work could focus on frameworks to propagate label uncertainty throughout the models; however, we believe this is outside the scope of this work, as this is the first paper to attempt fingerprinting with this type of data at this scale.

Attempts at computational speedups led to a striking result. Reducing the number of individuals in the training dataset has minimal impact on identification accuracy. This is likely because, in a one vs. rest paradigm, the negative class is ``walking patterns unlike the target individual.'' Beyond a few hundred individuals in the training data, adding more negative cases is redundant and does not increase model performance. Because runtime scales linearly with training data size, training on $n=500$ instead of $n=13367$ reduces computational burden by about 25-fold with no loss in accuracy.

For person identification using free-living accelerometry data, we recommend (1) using a computationally efficient method like \texttt{stepcount} to identify walking, (2) obtaining fingerprint-based predictors from areas of sustained walking, (3) using predictors from at least 30 minutes of walking data in weighted logistic regression models, and (4) training on randomly selected subsets of $\approx 500$ individuals instead of training on the full data to reduce computational burden. Despite the massive size of the data, we estimate this analysis – starting with walking labels in 7 full days of accelerometry from $n=15{,}000$ individuals and fitting models for subject identification – could be completed on a standard computer in 21 hours using 3 minutes of data per person, and 9 days if using 30 minutes of data per person, without the use of parallelization; running the logistic regression models in parallel would lead to substantial improvements.

Limitations of the approach include the computation time required to run ADEPT to identify walking (700 days on a standard computer). However, once walking segments have been obtained, computing the grid cell predictors and fitting logistic regression models can be easily parallelized and is computationally efficient. 

Future directions include investigating other methods of gait-based identification with this data, including deep learning, which has had promising results using other types of gait data \citep{deep_learning}, but also poses computational challenges, or treating the ``fingerprint'' as an image and using it as a predictor in scalar on image regression \citep{soir}. Other train/test paradigms could be examined, such as training on all data except the last day of observation, and testing on the last day. Finally, unlike deep learning methods, our method provides images (``fingerprints'') associated with each individual; another future direction includes clustering based on these images and investigating the association between images and health outcomes.

\begin{funding}
This work was supported by the National Institutes of Health under Grants R01NS060910 and R01AG075883. 
\end{funding}

\begin{supplement}
\noindent \textbf{Code}\\
All R code for the analysis available on Github at \url{https://github.com/lilykoff/nhanes_fingerprinting}. Additionally, code for walking recognition and obtaining the grid-cell predictors from raw accelerometry is available as an R package \texttt{accelPrint}, available for download from \url{https://github.com/lilykoff/accelPrint} or in R using the command \\ \texttt{pak::pak(``lilykoff/accelPrint'')}

\noindent \textbf{Data}\\
Raw data are available to download from NHANES at \url{https://wwwn.cdc.gov/Nchs/Data/Nhanes/Public/2011/DataFiles/PAX80_G.htm} and \url{https://wwwn.cdc.gov/Nchs/Data/Nhanes/Public/2013/DataFiles/PAX80_H.htm}
\end{supplement}


\bibliographystyle{imsart-nameyear} 
\bibliography{references}       


\end{document}